\documentclass[fleqn,usenatbib]{mnras}

\usepackage{newtxtext,newtxmath}
\usepackage{amsmath}
\usepackage{color}

\usepackage[T1]{fontenc}
\usepackage{threeparttable}
\usepackage[figuresright]{rotating}
\DeclareRobustCommand{\VAN}[3]{#2}
\let\VANthebibliography\thebibliography
\def\thebibliography{\DeclareRobustCommand{\VAN}[3]{##3}\VANthebibliography}

\usepackage{graphicx}	
\usepackage{amsmath}	

\title[Timing analysis of EXO 2030+375]{Timing analysis of EXO 2030+375 during its 2021 giant outburst observed with \textit{Insight}-HXMT}

\author[Y. C. Fu et al.]{
Yu-Cong Fu,$^{1,2,3}$\thanks{E-mail: fuyucong@xao.ac.cn}
L. M. Song,$^{2,3}$\thanks{E-mail: songlm@ihep.ac.cn}
G. Q. Ding,$^{1}$\thanks{E-mail: dinggq@xao.ac.cn}
M. Y. Ge,$^{3}$
Y. L. Tuo,$^{3}$
S. Zhang,$^{3}$
S. N. Zhang,$^{3}$
X. Hou,$^{4,5}$
\newauthor
J. L. Qu,$^{3}$
J. Zhang,$^{3}$
L. Zhang,$^{3}$
Q. C. Bu,$^{6}$
Y. Huang,$^{3}$
X. Ma,$^{3}$
X. Zhou,$^{1,7,8}$
W. M. Yan,$^{1,7,8}$
Z. X. Yang,$^{2,3}$
\newauthor
X. F. Lu,$^{2,3}$
T. M. Li,$^{2,3}$
Y. C. Xu,$^{2,3}$
P. J. Wang,$^{2,3}$
S. H. Xiao,$^{1,2,3}$
H. X. Liu,$^{2,3}$
X. Q. Ren,$^{2,3}$
Y. F. Du,$^{2,3}$
\newauthor
Q. X. Zhao,$^{2,3,4}$
and Y. X. Xiao$^{2,3}$
\\
$^{1}$Xinjiang Astronomical Observatory, Chinese Academy of Sciences, Urumqi, Xinjiang 830011, China\\
$^{2}$University of Chinese Academy of Sciences, Chinese Academy of Sciences, Beijing 100049, China\\
$^{3}$Key Laboratory of Particle Astrophysics, Institute of High Energy Physics, Chinese Academy of Sciences, Beijing 100049, China\\
$^{4}$Yunnan Observatories, Chinese Academy of Sciences, Kunming 650216, China\\
$^{5}$Key Laboratory for the Structure and Evolution of Celestial Objects, Chinese Academy of Sciences, Kunming 650216, China\\
$^{6}$Institut für Astronomie und Astrophysik, Kepler Center for Astro and Particle Physics, Eberhard Karls Universität, Sand 1, 72076 Tübingen, Germany\\
$^{7}$Key Laboratory of Radio Astronomy, Chinese Academy of Sciences, Nanjing 210008, China\\
$^{8}$Xinjiang Key Laboratory of Radio Astrophysics, Urumqi 830011, China
}

\date{Accepted XXX. Received YYY; in original form ZZZ}

\pubyear{2022}

\begin{document}
\label{firstpage}
\pagerange{\pageref{firstpage}--\pageref{lastpage}}
\maketitle

\begin{abstract}
We report the evolution of the X-ray pulsations of EXO 2030+375 during its 2021 outburst using the observations from \textit{Insight}-HXMT. 
Based on the accretion torque model, we study the correlation between the spin frequency derivatives and the luminosity.
Pulsations can be detected in the energy band of 1--160 keV.
The pulse profile evolves significantly with luminosity during the outburst, leading to that the whole outburst can be divided into several parts with different characteristics.
The evolution of the pulse profile reveals the transition between the super-critical (fan-beam dominated) and the sub-critical accretion (pencil-beam dominated) mode.
From the accretion torque model and the critical luminosity model, based on a distance of 7.1 kpc, the inferred magnetic fields are $(0.41-0.74) \times 10^{12}$ G and $(3.48-3.96) \times 10^{12}$ G, respectively, or based on a distance of 3.6 kpc, the estimated magnetic fields are $(2.4-4.3) \times 10^{13}$ G and $(0.98-1.11)\times 10^{12}$ G, respectively. Two different sets of magnetic fields both support the presence of multipole magnetic fields of the NS.
\end{abstract}



\section{Introduction} \label{sec:intro}

A Be/X-ray binary (BeXRB) system consists of a neutron star (NS) and a Be stellar companion, and BeXRBs are among the brightest X-ray sources. 
The X-ray emission from a BeXRB is due to the accretion from the circumstellar disk onto the NS.
Meanwhile, the angular momentum carried by the accretion flow is also transferred to the NS.
Thus, the properties of emission during an outburst provide a physical correlation between the spin-up rate and the accretion rate \citep[e.g.,][]{1979ApJ...234..296G, Wang_1996, 2019ApJ...879...61Z, 2020JHEAp..27...38T}.
BeXRB transient binaries exhibit two types of typical outbursts, i.e. type-I (normal) outbursts and type-II (giant) outbursts. Type-I outbursts are characterized by lower X-ray luminosity of $\textit{L}_{\rm{X}} \sim10^{36} \ \rm {erg \ s^{-1}}$ and associated with the orbital period cycle, while type-II outbursts are characterized by high X-ray luminosity of $\textit{L}_{\rm{X}} \gtrsim10^{37} \ \rm {erg \ s^{-1}}$ and generally last from several weeks to months \citep{1986ApJ...308..669S, 1997ApJS..113..367B, 2018ApJ...863....9W, 2020MNRAS.491.1851J}. 

The transient accreting pulsar EXO 2030+375 is a BeXRB, with a B0 Ve star as the optical companion \citep{1988MNRAS.232..865C}. 
In the past nearly forty years, the type-I outbursts of this source have been found during almost every periastron passage and the characteristics of them have been analyzed well. 
The X-ray pulsations were detected with a NS spin period of $\sim 42$ s \citep{1989ApJ...338..359P} and the orbital period was determined at $\sim 46$ days \citep{2008ApJ...678.1263W}.
The distance was estimated as $7.1\pm0.2$ kpc from the optical extinction \citep{2002ApJ...570..287W}, which has been adopted in most previous studies. However, \citet{2021MNRAS.502.5455A} updated the distance to $3.6^{+0.9}_{-1.3}$ kpc using \textit{Gaia}. The difference between the two values will significantly change the magnetic field measurements, so based on the two values of distance, the discussion will be given separately.

Since it was discovered by \textit{EXOSAT} in 1985 \citep{1985IAUC.4066....1P}, EXO 2030+375 had showed three giant outbursts in 1985, 2006, and 2021, respectively. During the 1985 giant outburst, the X-ray luminosity of the source reached $\textit{L}_{1-20 \ \rm{keV}} \sim 2 \times 10^{38} \ \rm {erg \ s^{-1}}$, and the spin-up time scale was determined at $-P/\dot{P}\sim 30 \ \rm{yr}$ \citep{1989ApJ...338..359P}. During the 2006 giant outburst, the X-ray luminosity of the source reached $\textit{L}_{1-20 \ \rm{keV}} \sim 1.2 \times 10^{38} \ \rm {erg \ s^{-1}}$, and the spin-up time scale was confirmed at $-P/\dot{P}\sim 40 \ \rm{yr}$ \citep{2007A&A...464L..45K}. In July 2021, \textit{MAXI/GSC} triggered on an X-ray activity from EXO 2030+375 \citep{2021ATel14809....1N}, and the source encountered the third giant outburst. The third outburst was weaker than the previous two outbursts, with a peak flux of ~550 mCrab \citep{2021ATel15006....1T}.The \textit{NICER} started monitoring on 28 July 2021 during the rise of the outburst \citep{2021ATel14911....1T}.
The X-ray luminosity of the source reached $\textit{L}_{0.7-10 \ \rm{keV}} \sim 0.4 \times 10^{38} \ \rm {erg \ s^{-1}}$ from the analysis of \textit{NICER} data, and the spin-up time scale was inferred at $-P/\dot{P}\sim 60 \ \rm{yr}$ from the analysis of \textit{Fermi}/GBM and \textit{Swift}/BAT data \citep{2022MNRAS.515.5407T}.  

In this work, using the data of \textit{Insight}-HXMT for EXO 2030+375 during its 2021 giant outburst, we perform the timing analysis of this source. The observations, as well as the reduction of \textit{Insight}-HXMT data, are presented in Section \ref{sec:OBSERVATIONS AND DATA REDUCTION}, the data analysis and results are described in Section \ref{sec:ANALYSIS AND RESULTS}, and, finally, the discussions and conclusions are given in Section \ref{sec:Discussion}.

\section{OBSERVATIONS AND DATA REDUCTION} \label{sec:OBSERVATIONS AND DATA REDUCTION}
\begin{table}
\centering
\caption{\textit{Insight}-HXMT observations of EXO 2030+375.}
\label{tab:table1}
\begin{tabular}{cccccc}
\hline
 {Obs.$^{\rm a}$} &  {Date} & {Obs. Time$^{\rm b}$}& 
 {Obs.$^{\rm a}$} &  {Date} & {Obs. Time$^{\rm b}$}\\
 { } &  {(MJD)} & {(s)}&  
 { } &  {(MJD)} & {(s)} \\
 \hline
    001   & 59423.20  & 18001  & 036 & 59476.08  & 35123  \\
    002   & 59425.13  & 17357  & 037 & 59477.37  & 35084  \\
    003   & 59427.30  & 17360  & 038 & 59479.23  & 23190  \\
    004   & 59429.32  & 35263  & 039 & 59481.34  & 35078  \\
    005   & 59430.88  & 35249  & 040 & 59483.52  & 35079  \\
    006   & 59431.74  & 35258  & 041 & 59486.33  & 34903  \\
    007   & 59432.73  & 35102  & 042 & 59488.52  & 23500  \\
    008   & 59433.73  & 34556  & 043 & 59489.68  & 40634  \\
    009   & 59435.06  & 35260  & 044 & 59491.64  & 34718  \\
    010   & 59436.05  & 35319  & 045 & 59493.51  & 34858  \\
    011   & 59437.11  & 35263  & 046 & 59495.43  & 34802  \\
    012   & 59438.12  & 35361  & 047 & 59497.43  & 35244  \\
    013   & 59439.18  & 35310  & 048 & 59499.13  & 34735  \\
    016   & 59442.03  & 35183  & 049 & 59501.40  & 35325  \\
    017   & 59443.82  & 35252  & 050 & 59503.51  & 34509  \\
    018   & 59445.68  & 35237  & 051 & 59506.42  & 35205  \\
    019   & 59447.85  & 35212  & 052 & 59507.81  & 34661  \\
    020   & 59450.86  & 35210  & 053 & 59509.53  & 34661  \\
    021   & 59451.95  & 40930  & 054 & 59511.38  & 34661  \\
    022   & 59452.98  & 35215  & 055 & 59513.40  & 40389  \\
    023   & 59453.71  & 35202  & 056 & 59515.61  & 35159  \\
    024   & 59455.00  & 35189  & 057 & 59517.43  & 34524  \\
    025   & 59458.24  & 35016  & 058 & 59519.48  & 34369  \\
    026   & 59461.91  & 35220  & 059 & 59521.51  & 40651  \\
    027   & 59464.13  & 63844  & 060 & 59523.59  & 46390  \\
    028   & 59464.86  & 29488  & 061 & 59525.54  & 40637  \\
    029   & 59466.12  & 58116  & 062 & 59527.43  & 35262  \\
    030   & 59467.14  & 29480  & 063 & 59529.38  & 41030  \\
    031   & 59468.20  & 35157  & 064 & 59531.40  & 34503  \\
    032   & 59469.03  & 34666  & 065 & 59533.45  & 34332  \\
    033   & 59470.12  & 35147  & 066 & 59535.37  & 34703  \\
    034   & 59471.18  & 40385  & 067 & 59539.34  & 35510  \\
    035   & 59473.43  & 35126  &     &           &        \\
\hline
\end{tabular}
\begin{tablenotes}
\item[]$^{\rm a}$ Observation ID, 001: P0304030NNN, NNN=001.\\
$^{\rm b}$ The total duration of the observation on the target source, not the good time interval after filtering.
\end{tablenotes}
\end{table}

EXO 2030+375 was observed by \textit{Insight}-HXMT from 2021 July 28 (MJD 59423) to 2021 November 21 (59539). There are 65 observations of the core proposal P0304030 with a total of ~2292 ks exposure time. Details of the observation info are presented in Table \ref{tab:table1}.

\textit{Insight}-HXMT \citep{2020SCPMA..6349502Z}, the first Chinese X-ray astronomy mission, consists of three main payloads: the High Energy X-ray Telescope (HE /20--250 keV, \citealp{2020SCPMA..6349503L}), the Medium Energy X-ray Telescope (ME /5--30 keV, \citealp{2020SCPMA..6349504C}) and the Low Energy X-ray Telescope (LE /1--15 keV, \citealp{2020SCPMA..6349505C}). The time resolution of the HE, ME, and LE instruments are $\sim25\,\mu$s, $\sim276\,\mu$s, and $\sim1$ ms, respectively.
The \textit{Insight}-HXMT provides continuous observations of EXO 2030+375, which can be used to investigate the timing and spectral properties of this source
\begin{figure}
    \includegraphics[width=\columnwidth]{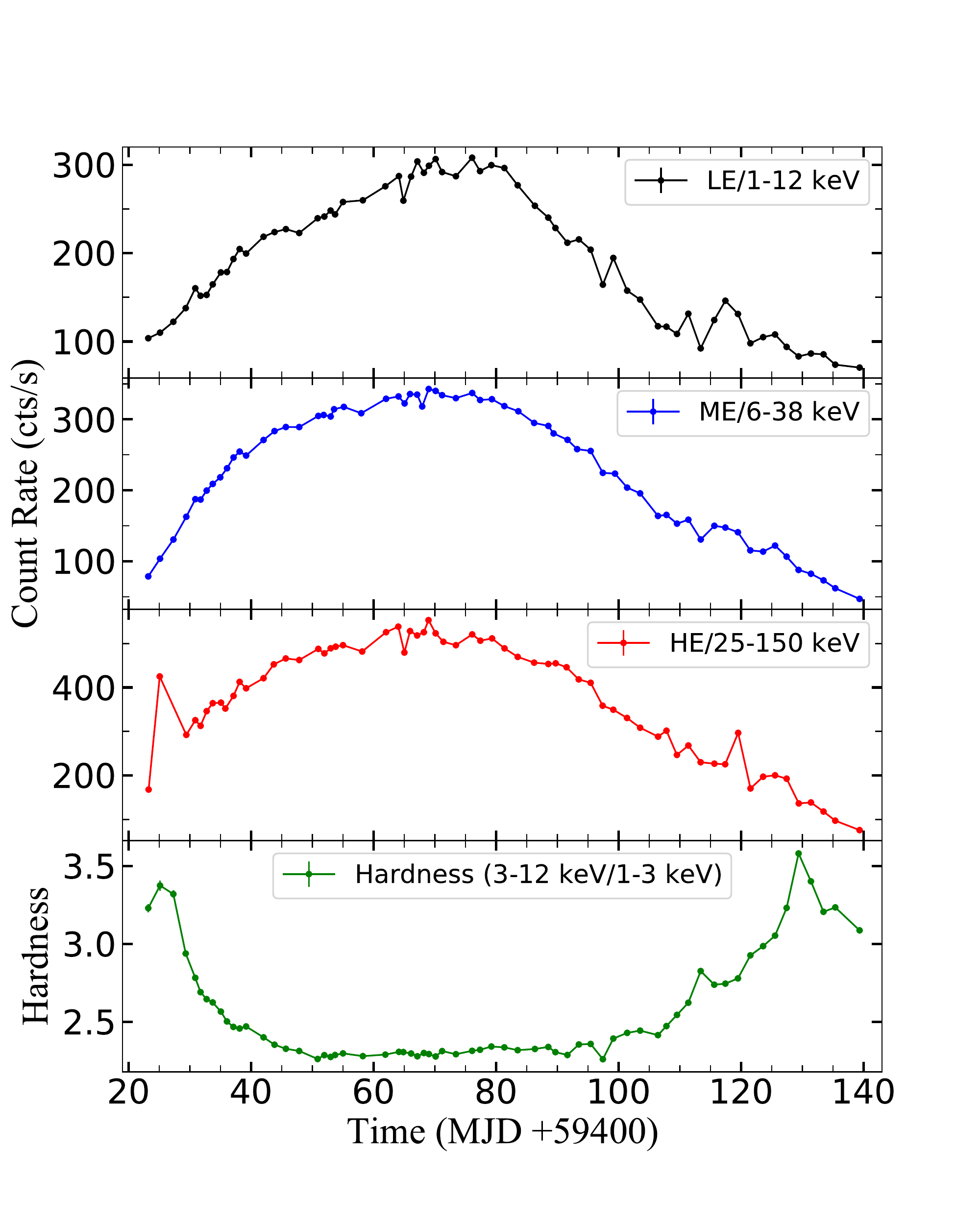}
    \caption{\textit{Insight}-HXMT net light curves and hardness of EXO 2030+375 of all observations from 2021 July 28 (MJD 59423) to 2021 November 21 (59539). Hardness is defined as the ratio of net count rate between the energy bands of 3--12 keV and 1--3 keV for \textit{Insight}-HXMT/LE data. Each point corresponds to one observation.}
    \label{fig:obs}
\end{figure}

The \textit{Insight}-HXMT Data Analysis Software\footnote{http://hxmtweb.ihep.ac.cn/software.jhtml} (\texttt{HXMTDAS}) v2.04 and \texttt{HXMTCALDB} v2.05 are used to analyze the raw data.
The pipeline of data reduction for \textit{Insight}-HXMT was introduced in previous publications \citep[e.g.,][]{2018ApJ...866..122H, 2018ApJ...864L..30C, 2021ApJ...919...92B, 2022RAA....22k5002F}.
We filter the data according to the following criteria for the selection of good time intervals (GTIs):
(1) elevation angle (ELV) $>$ 10$^{\circ}$;
(2) the value of the geomagnetic cutoff rigidity (COR) $>$ 8 GeV;
(3) elevation angle above bright earth for LE detector $>$ 30$^{\circ}$;
(4) the time before and after the South Atlantic Anomaly passage $>$ 100 s;
(5) the offset angle from the pointing direction $<$ 0.1$^{\circ}$.
Only small field of views (FoVs) are applied to avoid possible interference from the bright earth and local particles.
The background estimations based on the emission detected by blind detectors of the three instruments are performed with the Python scripts LEBKGMAP \citep{2020JHEAp..27...24L}, MEBKGMAP \citep{2020JHEAp..27...44G} and HEBKGMAP \citep{2020JHEAp..27...14L}, respectively.

The arrival times of photons are corrected to the solar system barycenter using the \texttt{HXMTDAS} task \texttt{hxbary}.
The events after the correction of the binary modulation are folded to obtain the pulse profile.
The background counts are far less than the counts from the source, and there is no pulse in the background, so the background does not affect the pulse profile, thus the events without background subtraction are used to generate the pulse profile.

In Figure \ref{fig:obs}, the evolution of the net count rate after data reduction for the three instruments are shown in the top three panels and the hardness is shown in the bottom panel.
The \textit{Insight}-HXMT observations cover the whole giant outburst.
\section{ANALYSIS AND RESULTS} \label{sec:ANALYSIS AND RESULTS}
\subsection{Spectral analysis}
\label{sec:Luminosity}
\begin{figure}
	\includegraphics[width=\columnwidth]{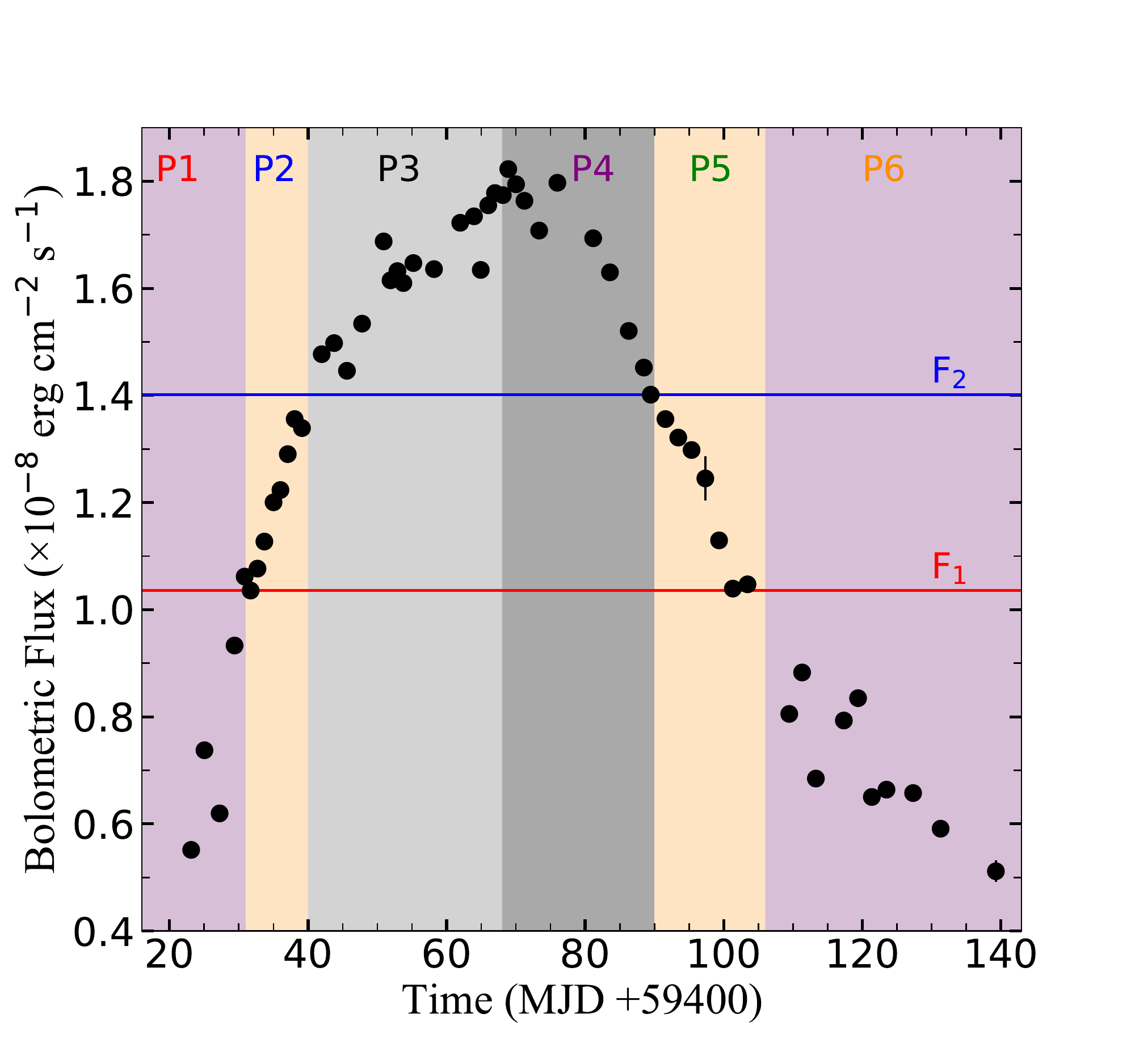}
    \caption{The evolution of flux (1--150 keV) is estimated by fitting the \textit{Insight}-HXMT spectra. The entire outburst is divided into six parts marked with P1--P6 to analyze the average pulse profile with a higher significance. $F_1=1.03\times 10^{-8} \ \rm {ergs \ cm^{-2} \ s^{-1}}$ and $F_2=1.40\times 10^{-8} \ \rm {ergs \ cm^{-2} \ s^{-1}}$ are the critical fluxes at the junctions of different parts.}
    \label{fig:flux}
\end{figure}
For all the \textit{Insight}-HXMT observations under consideration
for analysis, we obtain the fluxes and the luminosities.
The fluxes are estimated from the fitting of the broadband \textit{Insight}-HXMT spectra.
The spectra of EXO 2030+375 are dominated by the continuum emission and can be fitted by a simple power-law or cutoff power-law model without considering the absorption and emission features, which contribute to a negligible flux \citep{2007A&A...464L..45K, 2015RAA....15..537N, 2022MNRAS.515.5407T}.
The specific processes are as follows:
(1) Generating spectra, backgrounds, and response files;
(2) With \texttt{XSPEC} v12.11.1 \citep{1996ASPC..101...17A}, fitting each spectrum in 1--150 keV with the model of \texttt{TBabs*cutoffpl}\footnote{https://heasarc.gsfc.nasa.gov/docs/software/lheasoft/} \citep{2000ApJ...542..914W};
(3) Freezing the best-fitted norm of the model;
(4) Adding \texttt{cflux} component to the model and then using the model of \texttt{TBabs*cflux*cutoffpl} to calculate the unabsorbed flux of \texttt{cutoffpl}.
The obtained values of photon index ($\Gamma$) and e-folding energy ($E_{\rm{cut}}$) are in the ranges of 0.79--1.73 and 14--34 keV, respectively. The values of reduced chi-squared ($\chi^2_{\nu}$) for every observation are less than 1.16 with 1360 degrees of freedom. The detailed spectral analysis with the \textit{Insight}-HXMT data for this source is ongoing and the results will be published in another paper.

The variation of the flux during the outburst has been shown in Figure \ref{fig:flux}.
Based on the distance of 7.1 kpc \citep{2002ApJ...570..287W}, the luminosity increases from $0.33\times 10^{38} \ \rm {erg \ s^{-1}}$ on MJD 59523.20 to the maximum value of $1.10\times 10^{38} \ \rm {erg \ s^{-1}}$ on MJD 59468.20, and then falls back to $0.31\times 10^{38} \ \rm {erg \ s^{-1}}$ on MJD 59539.34. 
Based on the distance of 3.6 kpc \citep{2021MNRAS.502.5455A}, the luminosity increases from $0.85\times 10^{37} \ \rm {erg \ s^{-1}}$ on MJD 59523.20 to the maximum value of $2.82\times 10^{37} \ \rm {erg \ s^{-1}}$ on MJD 59468.20, and then falls back to $0.79\times 10^{37} \ \rm {erg \ s^{-1}}$ on MJD 59539.34.

According to the variation of pulse profile, the entire outburst is divided into six parts as shown in Figure \ref{fig:flux} (see description in Section \ref{sec:Pulse profile}). 
P1--P3 are in the rising parts of the outburst, and P4--P6 are in the falling parts.
At the junction of different parts, the two critical fluxes are defined as $F_1=1.03\times 10^{-8} \ \rm {ergs \ cm^{-2} \ s^{-1}}$ (red line) and $F_2=1.40\times 10^{-8} \ \rm {ergs \ cm^{-2} \ s^{-1}}$ (blue line).

\subsection{Evolution of spin frequency}
\label{sec:Spin frequency}
\begin{figure}
	\includegraphics[width=\columnwidth]{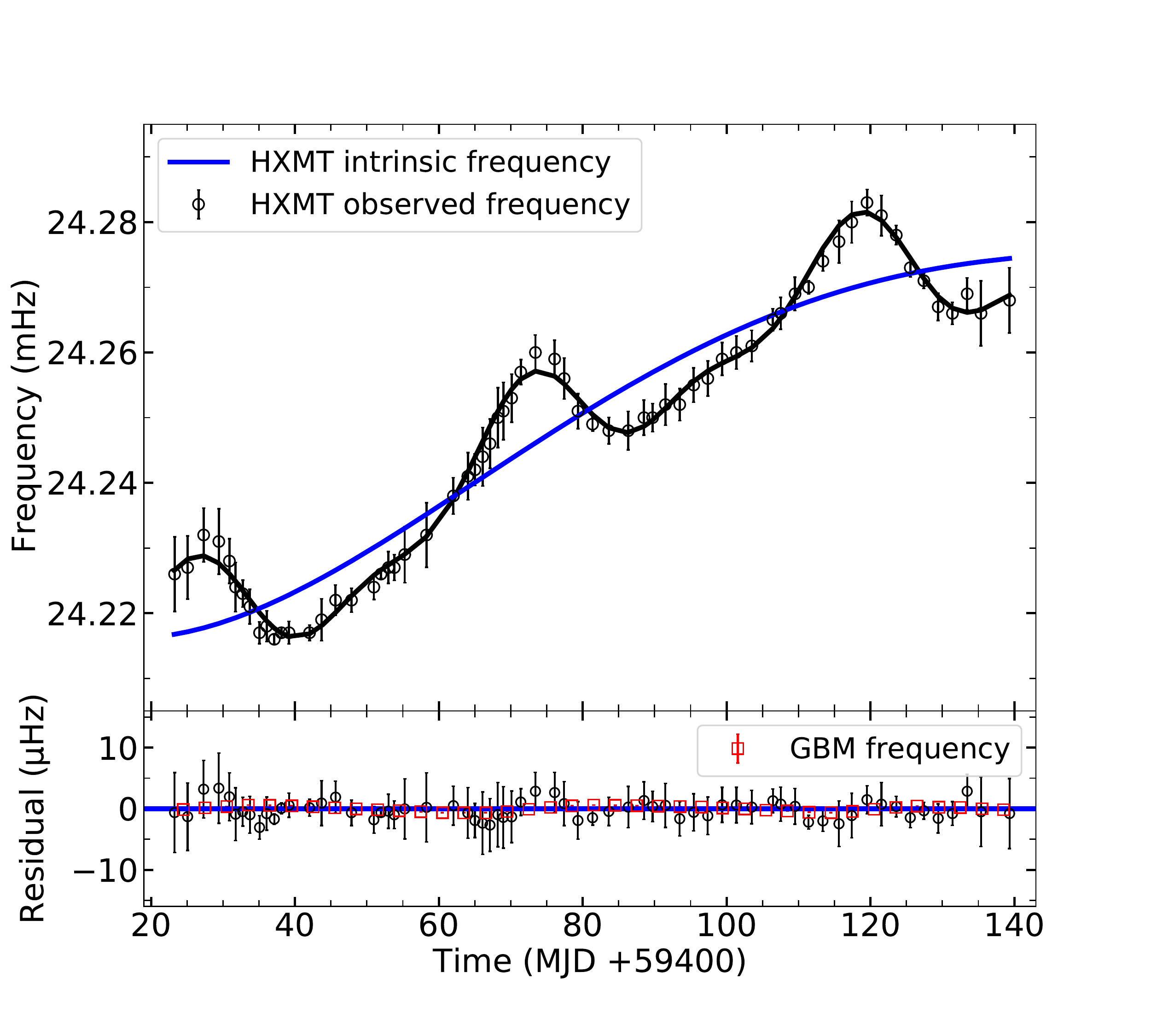}
    \caption{The spin frequencies of EXO 2030+375 obtained from \textit{Insight}-HXMT data are shown in the upper panel. The observed frequencies for \textit{Insight}-HXMT data are presented with black circles. The intrinsic frequencies for the \textit{Insight}-HXMT data are shown with the blue line. The residuals between the best-fit model and data are presented in the bottom panel. The frequencies for \textit{Fermi}/GBM data are shown with red squares for comparison.}
    \label{fig:spin}
\end{figure}
\begin{table}
\centering
\caption{Orbital and temporal parameters of EXO 2030+375 from \textit{Insight}-HXMT observations.}
\label{tab:table2}
\begin{tabular}{lc}
\hline
{Parameter} & {Result (Error)}\\
\hline
        $P_{\rm orb}$ (days)& 46.02217(35)  \\
        $e$& 0.4102(8) \\
	    $\omega$ (deg)&211.982(11) \\
	    $a_X$sin$i$ (lt s)&243.9(3) \\
	    $T_0$ (MJD)& 59423.20\\
	    $f_0$ (Hz)& 0.024217\\
	    $T_{\pi/2}$ (MJD)&52831.88(8)\\
	    $\dot f\ (\rm Hz\, s^{-1})$&2.5432(23)$\times 10^{-12}$\\
	    $\ddot f\ (\rm Hz\, s^{-2})$&3.227(5)$\times 10^{-13}$\\
	    $\dddot f\ (\rm Hz\, s^{-3})$&-1.038(11)$\times 10^{-14}$\\
	    $\ddddot f\ (\rm Hz\, s^{-4})$&1.19(13)$\times 10^{-16}$\\
	    \hline
	    $\chi^2/\rm dof$&199/169 \\
\hline
\end{tabular}
\end{table}

\begin{figure*}
	\includegraphics[width=\textwidth]{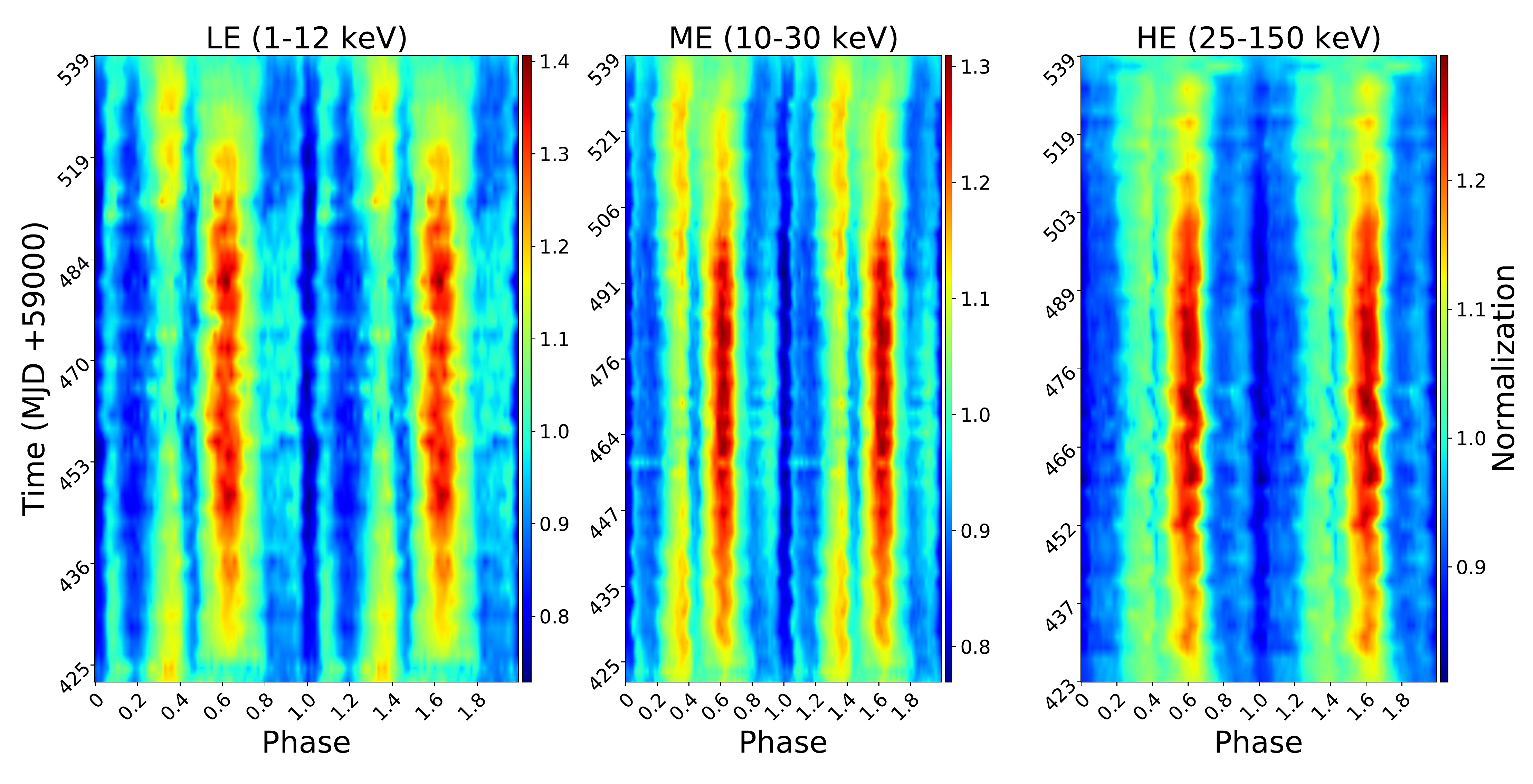}
    \caption{The two-dimensional (2D) maps describe the evolution of the pulse profile with time for three instruments LE (1--12 keV), ME (10--30 keV), and HE (25--150 keV). The colors representing the values of the pulse profile, which are normalized by Pulse/Average count rate. The main peak at $\sim$ 0.60 phase and the secondary peak at $\sim$ 0.35 phase are above the mean count rate, the minor peaks at $\sim$0.10 and $\sim$0.95 phase are close to or lower than the average count rate. 64 bins within a phase are used to generate the pulse profiles, and the plots are smoothed for clarity. Phase zero is defined as the minimum value of the pulse profile.}
    \label{fig:2D}
\end{figure*}
The observed spin frequency is calculated from each observation by using the epoch-folding technique \citep{1987A&A...180..275L}. 
Uncertainties for the spin frequency are estimated from the width of $\chi^2$ distribution for the trial periods.
However, the observed frequencies combine the intrinsic spin frequency of the NS and the effect of the Doppler shift due to the binary motion.
To obtain the intrinsic spin frequency of the pulsar, the orbital motion of the binary must be corrected \citep[e.g.,][]{2012MNRAS.423.2854L,2017ApJ...843...69W}. 
The method described in \citet{2005ApJ...635.1217G} is applied to fit the evolution of spin frequency.
We use the orbital parameters of EXO 2030+375 from \citet{2008ApJ...678.1263W} as the starting values of our fitting with the \textit{Insight}-HXMT results.

The observed spin frequencies could be written as
\begin{equation}
    f(t)=f_{\rm{spin}}(t)-\frac{2\pi f_{a}\emph{a}_{\rm{X}}\rm{sin}\emph{i}}{P_{\rm{orb}}}(\rm{cos} \, \emph{l}+\emph{g} \, \rm{sin} \, 2\emph{l}+\emph{h} \, \rm{cos} \, 2\emph{l}),
    \label{eq:q1}
\end{equation}
where $f_{\rm{spin}}(t)$ is the time-dependent NS intrinsic spin frequency, $f_{a}$ is a constant approximating $f_{\rm{spin}}(t)$, $\emph{a}_{\rm{X}}\rm{sin}\emph{i}$ is the projected
orbital semi-major axis in units of light-travel seconds, $\emph{i}$ is the system inclination, and $P_{\rm{orb}}$ (days) is the orbital period. 
The coefficients $\emph{g}=\emph{e}\, \rm{sin}\,\omega$ and $\emph{h}=\emph{e}\, \rm{cos}\,\omega$ are the functions of eccentricity $\emph{e}$ and the longitude of periastron $\omega$. 
And $l=2\pi(t-T_{(\pi/2)})/P_{\rm{orb}}+\pi/2$ is the mean longitude, where the $T_{(\pi/2)}$ is the epoch when the mean longitude $l=\pi/2$.

The intrinsic spin frequency evolution is described by a fourth-order polynomial function,
\begin{equation}\label{eq:q2}
\begin{split}
    f_{\rm{spin}}(t)=f_0&+\dot{f}(t-t_0)+\frac{1}{2}\ddot{f}(t-t_0)^2\\
    &+\frac{1}{6}\dddot{f}(t-t_0)^3+\frac{1}{24}\ddddot{f}(t-t_0)^4,
\end{split}
\end{equation}
where $f_0$ is the frequency at the reference time $t_0$ of the first frequency measurement, $\dot{f}$, $\ddot{f}$, $\dddot{f}$, and $\ddddot{f}$ are the first, second, third, and fourth derivatives of the intrinsic spin frequency, respectively. 

We fit the \textit{Insight}-HXMT results with Equation \ref{eq:q1} and show the evolution of the observed spin frequency with the black circles in Figure \ref{fig:spin}.
After correcting the Doppler modulation due to the binary motion, we get the evolution of the intrinsic spin frequency as shown by the blue solid line.
The intrinsic spin frequency evolves from 24.217 mHz on MJD 59423.20 to 24.274 mHz on MJD 59539.34 with an average spin derivative of $5.75\times 10^{-12} \ \rm{Hz\,s^{-1}}$.
The bottom panel shows the residuals between the fitting model and the data.
In addition, the \textit{Fermi}/GBM spin frequencies are also shown here with the red squares for comparison.
The errors of the spin frequency of \textit{Insight}-HXMT results are about $10^{-6}$ Hz, which are larger than the statistical fluctuation of the data. 
The best-fit results are listed in Table \ref{tab:table2}, and the reduced chi-squared ($\chi^2_{\nu}$) is 1.18 for 169 degrees of freedom. The errors in parentheses are calculated with 1-$\sigma$ level uncertainties.

\subsection{Pulse profile}
\label{sec:Pulse profile}

For each observation, the obtained spin frequencies of the NS are used to produce the pulse profiles.
The profiles of different observations are aligned together using the cross-correlation function, and phase zero is defined as the minimum value of the pulse profile.
Then, all the profiles are plotted in a heatmap to show the evolution of the pulse profiles during the whole outburst. 
As shown in Figure \ref{fig:2D}, a double-peaked structure of the pulse profile appears in all the three instruments, and the phase of peaks remains unchanged during the outburst. 
Depending on the apparent difference in intensities, the peak on the right ($\sim$ 0.60 phase) is considered the main peak, and the peak on the left ($\sim$ 0.35 phase) is considered the secondary peak.
The intensity of the main peak evolves significantly, while the evolution of the secondary peak is more complex and has a different trend from the main peak.
Besides, at about 0.10 phase and 0.95 phase, there are two weak peaks in LE and ME, and their phases are almost unchanged. 

\begin{figure*}
	\includegraphics[width=\textwidth]{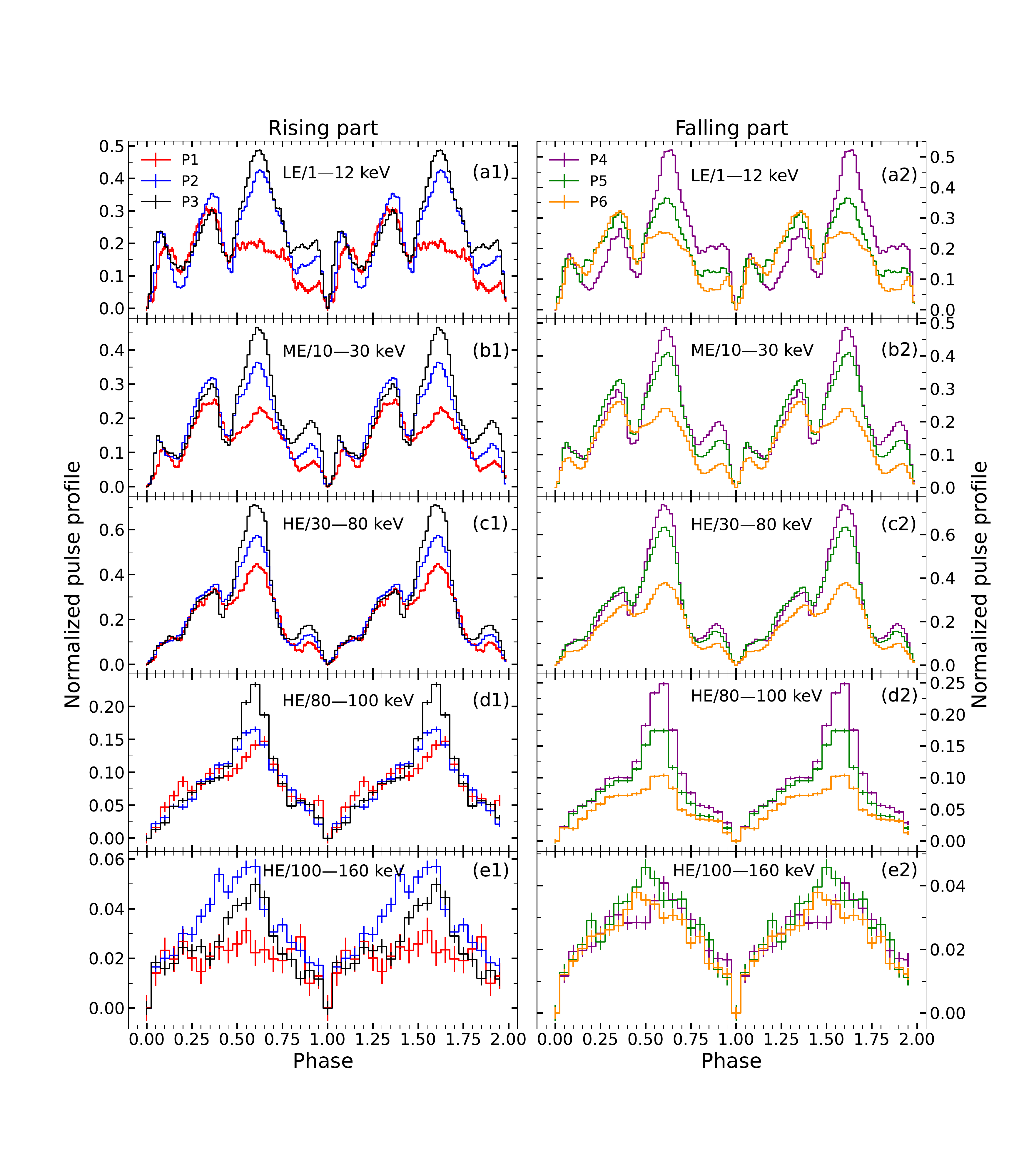}
    \caption{The evolution of the pulse profiles of EXO 2030+375 from \textit{Insight}-HXMT data. All the observations are split into six time intervals, which are described in detail in Section \ref{sec:Pulse profile}. Different parts are marked with different color symbols. The intensity of the pulse profiles are normalized by (Pulse - PulseMin)/(Average count rate), where PulseMin is the minimum value of the pulse profile. The left or right five panels from top to bottom show the results with energy bands of LE/1--12 keV, ME/10--30 keV, HE/30--80 keV, HE/80--100 keV, and HE/100--160 keV, respectively.}
    \label{fig:profile}
\end{figure*}

To study the evolution of the pulse profiles for a higher significance,
according to the relative magnitude of the intensities of the main and secondary peaks in LE as shown in Figure \ref{fig:2D}, the entire outburst is divided into six parts as follows:
MJD 59423--59431 (P1, the main peak is smaller than the secondary peak),
MJD 59431--59440 (P2, the main peak is close to or higher than the secondary peak),
MJD 59440--59468 (P3, the main peak is much higher than the secondary peak),
MJD 59468--59490 (P4, same as P3),
MJD 59490--59506 (P5, same as P2),
and MJD 59506--59540 (P6, same as P1).
For each part, the average pulse profiles obtained using the data from
all the three instruments of \textit{Insight}-HXMT are shown in Figure \ref{fig:profile}.

First of all, the shape of the profiles evolves with energy.
There are four peaks at LE (1--12 keV) and ME (10--30 keV) as shown in Panels (a1), (a2), (b1), and (b2) of Figure \ref{fig:profile}, among which the main peak at $\sim 0.60$ phase, the secondary peak at $\sim 0.35$ phase, and the others are minor peaks. 
In the hard X-ray energy band of 30--160 keV covered by HE, only the main peak is more significant.
The evolution of the pulse profiles can be identified in the harder energy band of 80--100 keV as shown in Panels (d1) and (d2). In the energy band of 100--160 keV, the shape of the pulse profiles is not significant and the evolution of the pulse profiles is not obvious, but pulsations can still be detected above 100 keV as shown in Panels (e1) and (e2).
Then, the profiles also depend on the luminosity.
The three parts of the rising and falling parts are symmetrical (P1$\sim$P6, P2$\sim$P5, P3$\sim$P4), and the flux at the boundaries are $F_1=1.03\times 10^{-8} \ \rm {erg \ cm^{-2} \ s^{-1}}$ and $F_2=1.40\times 10^{-8} \ \rm {erg \ cm^{-2} \ s^{-1}}$.
In all parts, we see a double-peaked shape (the main and secondary peaks) for LE and ME, but for HE, the secondary peak becomes weaker at 30--80 keV, while above 80 keV, only the main peak is prominent and the secondary peak is not visible. 
The evolution trend of the main peak is different from that of the secondary peak.
In the energy band below 80 keV, as shown in Panels (a1) to (c2), the main peak becomes stronger with the increase of luminosity, while the secondary peak has no obvious correlation with luminosity in both rising and falling parts.
It is worth noting that the intensity of the secondary peak is higher than that of the main peak during the part of lowest luminosity (P1 and P6, $F<F_1=1.03\times 10^{-8} \ \rm {erg \ cm^{-2} \ s^{-1}}$). 
As the luminosity increases, the main peak increases close to the secondary peak and then exceeds it (P2 and P5, $F_1<F<F_2$).
When the luminosity reaches its peak, the main peak is much higher than the secondary peak (P3 and P4, $F>F_2=1.40\times 10^{-8} \ \rm {erg \ cm^{-2} \ s^{-1}}$).

\begin{figure}
	\includegraphics[width=\columnwidth]{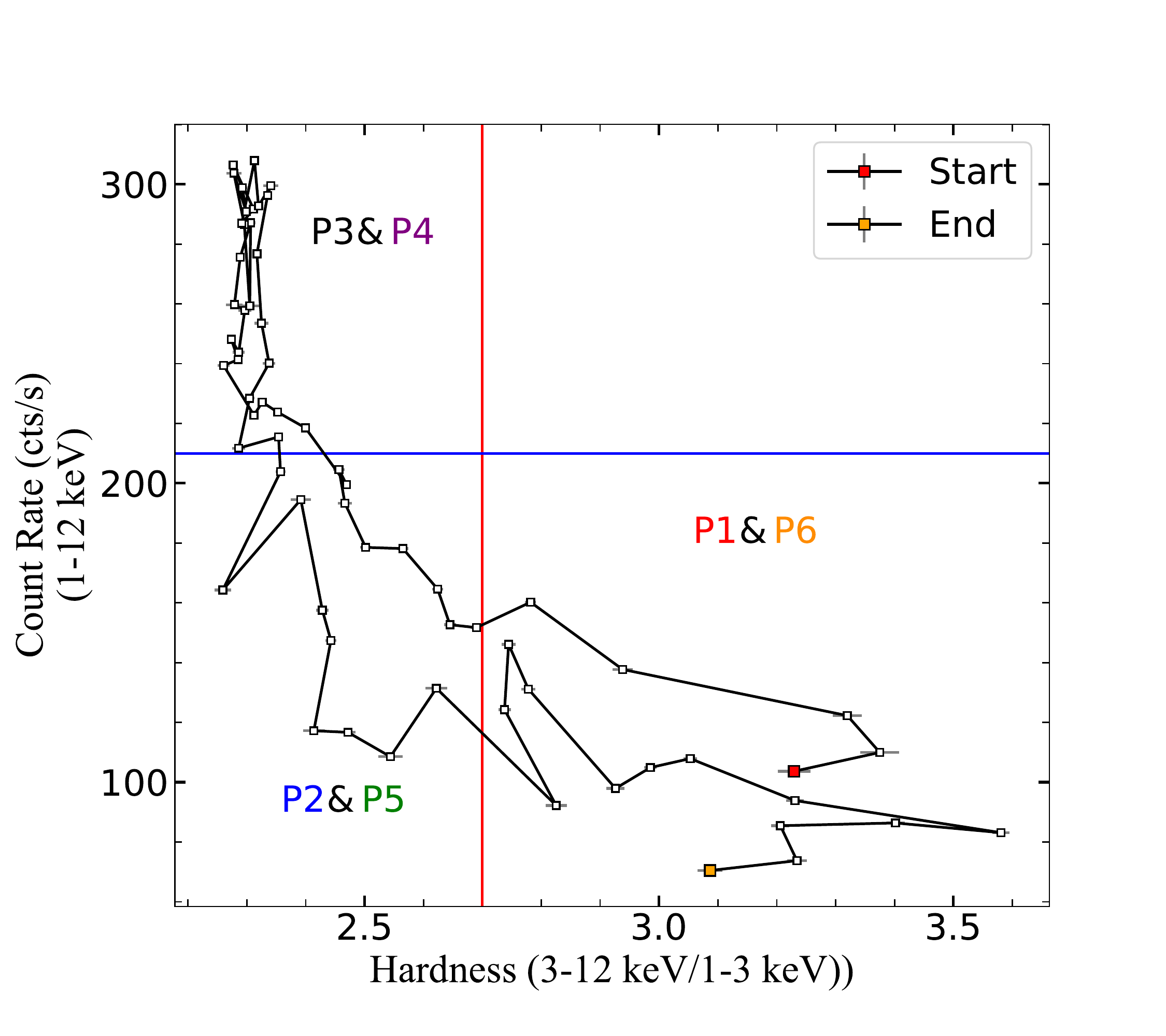}
    \caption{The hardness intensity diagram (HID) extracted from \textit{Insight}-HXMT/LE data. The hardness is defined as the count rate ratio  between 3--12 keV and 1--3 keV energy bands. 
    We use the observations in the rising part that correspond to a flux $F_1$ and $F_2$ in Figure \ref{fig:flux} to draw the two lines. 
    The red (Hardness = 2.7) and blue (Count rate = 210 cts/s) lines correspond to $F_1$ and $F_2$, respectively, and they divide the HID into three regions.    
    The lower right area corresponds to P1 and P6, where the main peak is smaller than the secondary peak; the lower left area corresponds to P2 and P5, where the main peak is close to or higher than the secondary peak; the upper left area corresponds to P3 and P4, where the main peak is obviously much higher than the secondary peak.}
    \label{fig:HID}
\end{figure}

In Figure \ref{fig:HID}, $F_1$ and $F_2$ divide the hardness intensity diagram (HID) into three regions. 
The lower right area corresponds to P1 and P6, where the main peak is smaller than the secondary peak, the count rate is kept at a low level, and the hardness is continuously reduced. The lower left area corresponds to P2 and P5, where the main peak is close to the secondary peak and then higher than it.
The upper left area corresponds to P3 and P4, where the main peak is much higher than the secondary peak, the count rate rises rapidly, and the hardness remains approximately unchanged at $\sim 2.3$.

\section{Discussion and conclusions}
\label{sec:Discussion}
In this work, using the observation data from \textit{Insight}-HXMT, we investigate the temporal evolution of the X-ray pulsations of EXO 2030+375 during its 2021 outburst. The variation of luminosity during the outburst is shown in Figure \ref{fig:flux}.
The obtained orbital parameters and the intrinsic spin frequency parameters are listed in Table \ref{tab:table2}. The evolutions of the pulse profiles with luminosity and energy are presented in Figures \ref{fig:2D} and \ref{fig:profile}, respectively. Next, we estimate the NS magnetic field strength with two different models, i.e. the accretion torque model and the critical luminosity model.

\subsection{Accretion torque model}
\label{sec:Accretion torque}
Based on the model of \citet{1979ApJ...234..296G}(GL model), the correlation between the spin frequency derivatives and the luminosity is used to investigate the accretion torque behavior during the outburst.
The spin evolution of X-ray pulsars driven by accretion torque during the outburst can be written as follow \citep{1977ApJ...217..578G}
\begin{equation}\label{eq:q3}
    -\dot{P}=\frac{NP^2}{2\pi I},
\end{equation}
where $N$ is the total torque, and $I$ is the effective moment of inertia of the NS.
The torque can be written by
\begin{equation}\label{eq:q4}
    N=n(\omega_{\rm{s}})\,N_0=n(\omega_{\rm{s}}) \, \dot{M}(GMr_0)^{1/2},
\end{equation}
where $n(\omega_{\rm{s}})$ is the dimensionless accretion torque, $\dot{M}$ is the mass accretion rate, $M$ is the mass of the NS, and $r_0$ is the magnetospheric radius.
From the GL model, the correlation between the spin frequency derivative ($\dot{f}$) of the pulsar and the X-ray luminosity can be written in the following form
\begin{equation}\label{eq:q5}
\begin{split}
    &\dot{f}=-\frac{\dot{P}}{P^2}\\&=5.0 \times 10^{-5} \mu_{30}^{2/7}n(\omega_{\rm{s}})R_{6}^{6/7}I_{45}^{-1}(\frac{M}{M_{\odot}})^{-3/7}L_{37}^{6/7}\rm{Hz\,yr^{-1}},
\end{split}
\end{equation}
where $\mu_{30}$ is the NS magnetic dipole moment $\mu$ ($\mu=\frac{1}{2}BR^3$) in the disk plane in units of $10^{30}\, \rm{G\,cm^{3}}$, $B$ is the magnetic field strength at the pole,
$L_{37}$ is the accretion luminosity in units of $10^{37}\,\rm{ergs\,s^{-1}}$, $R_6$ is the NS radius in units of $10^6$ cm, $I_{45}$ is the moment of inertia of the NS in units of $10^{45}\,\rm{g\,cm^2}$, and $\frac{M}{M_{\odot}}$ is the mass of the NS in units of the solar mass. The dimensionless torque $n(\omega_{\rm{s}})$ can be estimated as
\begin{equation}\label{eq:q6}
    n(\omega_{\rm{s}})\approx1.39\times \frac{1-\omega_{\rm{s}}[4.03(1-\omega_{\rm{s}})^{0.173}-0.878]}{1-\omega_{\rm{s}}},
\end{equation}
where $\omega_{\rm{s}}$ is the fastness parameter \citep{1977ApJ...215..897E}. For the slow rotator NS in EXO 2030+375 ($\omega_{\rm{s}}\ll1$), $n(\omega_{\rm{s}})\approx1.4$. 
\begin{figure}
	\includegraphics[width=\columnwidth]{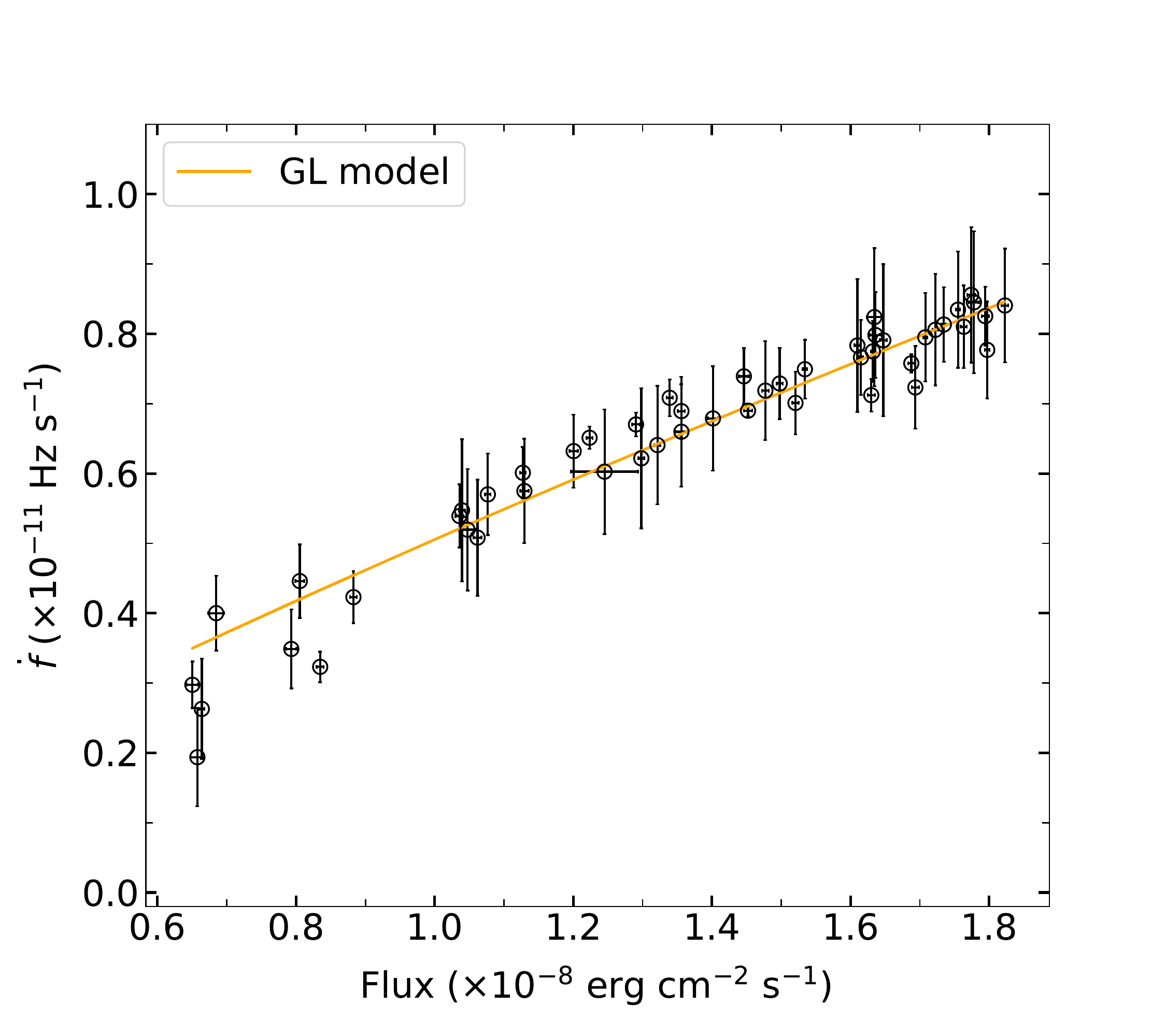}
    \caption{The correlation between the spin frequency derivatives and the flux (1--150 keV) observed by \textit{Insight}-HXMT. The orange solid line is fitted with the GL model.}
    \label{fig:GL}
\end{figure}
To analyze the accretion spin-up behavior of EXO 2030+375, we calculate the frequency derivatives $\dot{f}$ using \textit{Insight}-HXMT results after correcting the Doppler modulation. The $\dot{f}$ are obtained by $\Delta f /\Delta t$ for the time intervals between every two observations \citep{2018A&A...613A..19D,2020JHEAp..27...38T}, $\Delta t$ is the time interval between the two adjacent observations.
Since the corresponding time of $\dot{f}$ is the midpoint of $\Delta t$, which is inconsistent with the corresponding time of flux measurement, 
we interpolate the $\dot{f}$ using the linear interpolation method to match the times corresponding to through flux.
The errors of $\dot{f}$ are obtained from that of $f$ by the error propagation.
The correlation between $\dot{f}$ and the flux (1--150 keV) observed by \textit{Insight}-HXMT is shown in Figure \ref{fig:GL}. 
The $\dot{f}$ presents a positive correlation with the luminosity. 
The data points are well fitted with the GL model according to Equation \ref{eq:q5}. The fitting result reveals a correlation as follow 
\begin{equation}\label{eq:q7}
    D\approx 6.11\times B_{12}^{-1/6} \ [\rm{kpc}],
\end{equation}
where $D$ is the distance of the NS, $B_{12}$ is the magnetic field strength in units of $10^{12}$ G.
Considering the distance of $7.1\pm0.2$ kpc \citep{2002ApJ...570..287W}, the magnetic field is $ \sim 0.41 \times 10^{12}$ G. However, considering the distance of $3.6^{+0.9}_{-1.3}$ kpc \citep{2021MNRAS.502.5455A}, the magnetic field is $ \sim 2.4 \times 10^{13}$ G. Nevertheless, for a slow rotator, \citet{1995ApJ...449L.153W} model gave $n(\omega_{\rm{s}})=7/6$, which results in that the estimated magnetic field strength will be a factor of 1.8 larger than that inferred from the GL model. Thus, based on the two values of the distance, the dipole magnetic field strengths inferred from the different torque models are $(0.41-0.74) \times 10^{12}$ G (for 7.1 kpc) and $(2.4-4.3) \times 10^{13}$ G (for 3.6 kpc), respectively
\subsection{Critical luminosity and CRSF}
\label{sec:Pulse profiles and critical luminosity}
Combining the evolution of luminosity and hardness, we analyze the variability of the pulse profile and try to give the critical luminosity at the transition of the emission mode.
The evolution of the pulse profiles during the outburst can be explained in the context of the critical luminosity \citep{1975A&A....42..311B, 1976MNRAS.175..395B,2007ApJ...654..435B, 2012A&A...544A.123B, 2019MNRAS.489.1000W, 2020MNRAS.491.1851J, 2022ApJ...935..125W}. 
The emission mode transfers from the pencil-beam geometry at low luminosity level to the fan-beam emission geometry at high luminosity.
At lower luminosity, the deceleration of the accretion flow may occur via Coulomb breaking in a plasma cloud, the stopping region of the flow is just above the NS surface, and the emission from the stopping region escapes from the top of the column, forming a pencil beam \citep{1975A&A....42..311B, 1993ApJ...418..874N, 2012A&A...544A.123B}. 
At higher luminosity above $L_{\rm{crit}}$, the deceleration is dominated by the radiation pressure, and the emission primarily escapes through the column walls, forming a fan beam \citep{1976MNRAS.175..395B, 2012A&A...544A.123B, 2015MNRAS.447.1847M}.
The transition of the emission mode from the fan-beam to the pencil-beam geometry is usually accompanied by a transition of the pulse profile shape in the lower energy band, such as a transition from a double-peak pattern to a one-peak pattern \citep{2008ChA&A..32..241C}, which is observed in some sources.
For 4U 1901+03 \citep{2020JHEAp..27...38T}, the transition from the fan-beam to the pencil-beam is accompanied by a change in pulse profile from the double-peak to the one-peak patterns at 2--30 keV, while above 30 keV, the pulse profile remains a one-peak pattern. As for Swift J0243.6+6124 \citep{2018ApJ...863....9W}, the same change in pulse profile from the double-peak to the one-peak patterns  appeared in the energy range 0.2--12 keV (\textit{NICER}) and 12--100 keV (\textit{Fermi}/GBM). The critical luminosity at which the emission mode changes depends on the NS magnetic field strength and it can be written as follow \citep{2012A&A...544A.123B}
\begin{equation}\label{eq:q8}
    L_{\rm{crit}}=1.49\times10^{37}\emph{B}_{12}^{16/15}\,\rm{erg\,s^{-1}}.
\end{equation}

To study the fan-beamed X-ray emission for RX J0209.6--7427, \citet{2022arXiv220814785H} considered the dependence of emission properties on energy, luminosity, and emission geometry together. They demonstrated that the lower energy photons (e.g., 1--40 keV) can contribute to both the fan and pencil beam patterns, and the higher energy photons (e.g., from about 50 to above 130 keV) will preferentially escape in the fan beam pattern. Thus, the pulse profiles in the higher energy bands can be used to identify the fan beam pattern.
At the subcritical accretion, the main emission escapes in the pencil beam and thus the main peak at low energies will be significantly misaligned from the main peak of the high energy emission, which is confirmed by this study. As shown in panels (a1) and (c1) of Figure \ref{fig:profile} in this study, the main peak at $\sim$ 0.35 phase in panel (a1) is misaligned from that at $\sim$ 0.60 phase in Panel (c1) for the P1 part.  
Once in the supercritical regime, the main peak of the low energy emission will be aligned with that at the high energy.
Based on the analysis by \citet{2022arXiv220814785H}, we discuss the pulse profiles of EXO 2030+375 to determine the transition of the emission mode.

As shown in Figure \ref{fig:profile}, the pulse profile in the higher energy band above 80 keV has only one peak, which is considered to be contributed by the fan-beamed emission. 
In the lower energy band below 80 keV, this peak ($\sim$ 0.60 phase) still exists and the phase is consistent.
In addition, another peak appears at $\sim$ 0.35 phase.
The two peaks are consistent with the results of \citet{2022arXiv220814785H} that the lower energy photons can contribute to both the fan and pencil beam patterns, and thus the peak at $\sim$ 0.35 phase is considered to be contributed mostly by the pencil-beamed emission.
It is noted that in the energy bands 1--12 keV or 10--30 keV, the amplitude of the main peak and the secondary peak changes.
In parts P1 and P6, the amplitude of the peaks at $\sim$ 0.35 phase is greater than that at $\sim$ 0.60 phase, which indicates that the emission mode is dominated by the pencil beam. 
In other parts P2--P5, the amplitude of the peaks at $\sim$ 0.35 phase is smaller than that at $\sim$ 0.60 phase, which indicates that the emission mode is dominated by the fan beam. Therefore, the transition from the pencil beam to the fan beam occurs between P1 and P2, and the transition from the fan beam to the pencil beam occurs between P5 and P6. 
The flux corresponding to the critical luminosity is thus around $F_1=1.03\times 10^{-8} \ \rm {erg \ cm^{-2} \ s^{-1}}$.
Considering that the transition should occur between the two observations, this flux should be in a range of $(0.93-1.07)\times 10^{-8} \ \rm {erg \ cm^{-2} \ s^{-1}}$.
Taking this value into Equation \ref{eq:q8}, the correlation between the distance and the magnetic field strength can be written by
\begin{equation}\label{eq:q9}
    D=(3.41-3.65)\times B_{12}^{8/15} \ [\rm{kpc}].
\end{equation}
For the two different values of the source distance, the magnetic field strengths inferred from the critical luminosity model are $(3.48-3.96) \times 10^{12}$ G (for 7.1 kpc) and $(0.98-1.11)\times 10^{12}$ G (for 3.6 kpc), respectively.

The detection of cyclotron resonance scattering features (CRSFs) is the only way to directly and reliably measure the NS surface magnetic field strength \citep[e.g.,][]{2019JHEAp..23...29X, 2020ApJ...899L..19G, 2022ApJ...933L...3K}. 
However, there has not been significant detection of CRSFs for EXO 2030+375. The suspected absorbing structures were found at different energies. For simplicity, \citet{2022MNRAS.515.5407T} interpreted the absorption feature at 10.12 keV \citep{2008ApJ...678.1263W} found using the \textit{NuSTAR} data as a CRSF. If this does be a CRSF, the corresponding magnetic field is $\sim 1.13 \times10^{12}$ G.
\citet{2008A&A...491..833K} also reported an absorption structure at about 63 keV and interpreted it as the first harmonic of about 36 keV \citep{1999MNRAS.302..700R}.
The corresponding magnetic field strength will be $\sim 4.03 \times10^{12}$ G. In addition to using CRSF, \citet{2021JApA...42...33J} inferred the NS magnetic field strength of this source from the propeller effect and reported it to be in the range of $(3-15)\times10^{12}$ G. Moreover, \citet{2017MNRAS.472.3455E} constrained the magnetic field strength in the range of $(4-6)\times10^{12}$ G, inferred from the BW model \citep{2007ApJ...654..435B, 2009A&A...498..825F}. 

If the distance of EXO 2030+375 is 7.1 kpc \citep{2002ApJ...570..287W}, the magnetic field strength estimated from the torque models is $(0.41-0.74) \times 10^{12}$ G, which is close to the result inferred from the CRSF of 10.12 keV. However, the magnetic field strength inferred from the critical luminosity model is $(3.48-3.96) \times 10^{12}$ G, which is in approximate agreement with the results estimated from the possible CRSF at $\sim36$ keV, or the propeller effect, or the BW mode. 

We note that different magnetic field measurements have also been reported in other sources. From a CRSF at $\sim146$ keV, \citet{2022ApJ...933L...3K} estimated the NS surface magnetic field strength of Swift J0243.6+6124 and gave it to be $\sim1.6\times10^{13}$ G, and its  critical luminosity was also consistent with a strong NS surface magnetic field strength of $\sim10^{13}$ G \citep{2022MNRAS.512.5686L, 2020ApJ...902...18K}. However, the magnetic field strength of the NS given by the GL model is only $\sim6\times10^{12}$ G \citep{2019ApJ...879...61Z}. Moreover, \citet{2020MNRAS.491.1857D} estimated a dipole component of the magnetic field strength, being $\sim10^{12}$ G.
This difference is explained by the presence of multipole magnetic field components, which dominates the magnetic field in the vicinity of the surface of the NS.
For RX J0209.6--7427\citep{2022arXiv220814785H}, the magnetic fields given by the torque models and the critical luminosity method are $(4.8-8.6)\times10^{12}$ G and $(1.7-2.2)\times10^{13}$ G, respectively, taking into account of the uncertainties from the different torque models and $L_{\rm{crit}}$ estimation method. 
The two values are also interpreted as the dipole and multipole magnetic fields of the NS, as suggested for Swift J0243.6+6124 \citep{2022ApJ...933L...3K} and SMC X-3 \citep{2017A&A...605A..39T}. Similarly, we interpret the two values of the NS magnetic ﬁeld strength estimated for EXO 2030+375 in this work as the dipole magnetic field strength ($B_{12}\sim0.41-0.74$) and the multipole magnetic field strength ($B_{12}\sim3.48-3.96$), respectively.
Although there are great differences among different NS binary systems, the similarity of magnetic field measurements may support the existence of the multipole magnetic field components.

Alternatively, if the distance of EXO 2030+375 is 3.6 kpc \citep{2021MNRAS.502.5455A}, the magnetic field strength inferred by the critical luminosity method is $(0.98-1.11)\times 10^{12}$ G, which is consistent with the result given by the possible CRSF at 10.12 keV.
However, a larger magnetic field of $(2.4-4.3) \times 10^{13}$ G is obtained from the the torque models, which is an order of magnitude larger than that of about $10^{12}$ G for most accreting pulsars. It seems that the strength of the dipole magnetic field is larger than that of the multipole field, which is opposite to the results for the distance of 7.1 kpc.
The same phenomenon has also been discussed in other sources. For example, \citet{2020MNRAS.491.1851J} also presented the difference in the magnetic field strengths inferred between the GL model and the critical luminosity model for 2S 1417--624. If the distance from the optical measurement (9.9 kpc) is adopted, the magnetic field strength inferred from the GL model is much larger than that estimated from the critical luminosity model, as shown in Figure 4 of \citet{2020MNRAS.491.1851J}). They suggested that in addition to the uncertainty of the measurement method, the quadrupolar magnetic field might also be important.

For EXO 2030+375, the calculation of the magnetic field depends on the distance, which makes it necessary to have a reliable and solid measurement of distance. Both the different sets of magnetic field strength inferred with different values of distance support the presence of multipole magnetic fields of the NS.
However, we also could make a strong assumption that the dipole magnetic field dominates the NS surface magnetic field, and therefore the result of the torque model is the same as that of the critical luminosity model. If so, by simultaneously solving Equations \ref{eq:q7} and \ref{eq:q9}, the distance obtained is $(5.31-5.40)$ kpc, and the magnetic field strength of the NS is $(2.09-2.30) \times 10^{12}$ G. On the other hand, a solid detection of CRSFs in EXO 2030+375 would allow us to clarify the situation substantially. In the mean time, finding the similar phenomenon in more sources may also help us understand the topology of the magnetic fields of accreting NSs.

\section*{Acknowledgements}
We are grateful for the anonymous referee's constructive suggestions and comments.
This work has made use of data from the \textsl{Insight}-HXMT mission, a project funded by China National Space Administration (CNSA) and the Chinese Academy of Sciences (CAS), and data and software provided by the High Energy Astrophysics Science Archive Research Center (HEASARC), a service of the Astrophysics Science Division at NASA/GSFC. 
This work is supported by the National Key R\&D Program of China (Grant No. 2021YFA0718500), the Opening Foundation of Xinjiang Key Laboratory (No. 2021D0416), the Open Program of the Key Laboratory of Xinjiang Uygur Autonomous Region (No. 2020D04049), the National Natural Science Foundation of China (NSFC) under grants U1838108, U1838201, U1838202, 11733009, 11673023, 12273100, U1938102, U2038104, and U2031205, the CAS Pioneer Hundred Talent Program (grant No. Y8291130K2), and the Scientific and Technological innovation project of IHEP (grant No. Y7515570U1).
This work is also partially supported by International Partnership Program of Chinese Academy of Sciences (Grant No.113111KYSB20190020).

\section*{Data Availability}

The data analysed in this work are available from the following archives:
\begin{itemize}
    \item \textsl{Insight}-HXMT -- \href{http://hxmtweb.ihep.ac.cn/}{http://hxmtweb.ihep.ac.cn/}
    \item \textit{Fermi}--  \href{https://gammaray.nsstc.nasa.gov/gbm/science/pulsars/lightcurves/exo2030.html}{https://gammaray.nsstc.nasa.gov/gbm}
\end{itemize}



\bibliographystyle{mnras}
\bibliography{paper} 

\bsp	
\label{lastpage}
\end{document}